\documentclass[twocolumn]{revtex4-1}
\usepackage{graphicx}
\usepackage[dvipsnames]{xcolor}
\usepackage{color}
\usepackage{amsfonts}
\newcommand{\phiclub}{\phi^{\mbox{\fontsize{5}{0}$\clubsuit$}}}
\newcommand{\had}{\hat{a}^+}

\newcommand{\ha}{\hat{a}}
\newcommand{\hbd}{\hat{b}^+}
\newcommand{\hb}{\hat{b}}

\newcommand{\ket}[1]{\left| #1 \right\rangle}
\newcommand{\bra}[1]{\left\langle #1 \right|}

\newcommand{\expect}[1]{\left\langle#1\right\rangle}
\newcommand{\eea}{\end{eqnarray}}
\newcommand{\bea}{\begin{eqnarray}}
\newcommand{\ee}{\end{equation}}
\newcommand{\be}{\begin{equation}}

\begin{document}
\title{On Nonlinear Amplification: Improved Quantum Limits for Photon Counting} 

\author{Tzula B. Propp}
\author{S.J. van Enk}
\affiliation{Department of Physics and
Oregon Center for Optical, Molecular \& Quantum Sciences\\
University of Oregon, Eugene, OR 97403}

\begin{abstract}
We show that detection of single photons is not subject to the fundamental limitations that accompany quantum linear amplification of bosonic mode amplitudes, even though a photodetector does amplify a few-photon input signal to a macroscopic output signal. Alternative limits are derived for \emph{nonlinear} photon-number amplification schemes with optimistic implications for single-photon detection. Four commutator-preserving transformations are presented: one idealized (which is optimal) and three more realistic (less than optimal). Our description makes clear that nonlinear amplification takes place, in general, at a different frequency $\omega'$ than the frequency $\omega$ of the input photons. This can be exploited to suppress thermal noise even further up to a fundamental limit imposed by amplification into a single bosonic mode. A practical example that fits our description very well is electron-shelving. 
\end{abstract}

\maketitle
\section{Quantum amplification and noise}The fundamental relations between quantum noise and quantum amplification are most straightforwardly derived in the Heisenberg picture.
Thus, the standard way \cite{caves1982} to describe linear phase-preserving quantum amplification of a bosonic mode amplitude $a$ is through Caves' 
relation for the annihilation operator $\hat{a}$,
\bea\label{caves}
\ha_{{\rm out}}=\sqrt{G}\ha_{{\rm in}}+\sqrt{G-1}\hbd_{{\rm in}},
\eea
where $\hbd$ is the creation operator corresponding to an independent auxiliary bosonic mode $b$.
Here the input field amplitude of mode $a$ is amplified by a factor of $\sqrt{G}$, but there is a cost: extra noise arising from the additional mode $b$ \footnote{This could be a mode internal to the detector.}.
If this mode contains (thermal) excitations, mode $a$ after amplification will contain excitations, too, and their number is multiplied by $G-1$. Even if mode $b$ is in the vacuum state, it still adds noise \cite{caves1982}. It is clear that this extra noise is due to the additional creation operator term proportional to $\sqrt{G-1}$ in (\ref{caves}), but since that term is necessary so as to preserve the standard bosonic commutation relation $[\ha_{{\rm out}},\had_{{\rm out}}]=\openone$ this tradeoff between linear amplification and added noise is fundamental. Indeed, phase-preserving linear amplification in proposed number resolving platforms using superconducting qubits have noise bounded by the Caves limit \cite{metelmann2014,clerk2010}.

Recently, there has been some effort to describe  all parts of the photo detection process, including amplification \cite{yang2019},  
fully quantum mechanically \cite{young2018b,young2018,vanenk2017,dowling2018}. 
One conclusion that may be drawn from that research is that there is no severe amplification-driven tradeoff between efficiency and (thermally induced) dark counts. In particular, even though a few-photon signal must be amplified to a macroscopic level [forcing us to consider $G\gg 1$], thermal fluctuations in internal detector modes do not get amplified by the same factor of $G$. Experiments \cite{marsili2013,wollman2017uv}  on superconducting nanowires demonstrate that over a wide range of detected wavelengths dark count rates can indeed be extremely low (on the order of one dark count per day). How can we reconcile these results with that of the previous paragraph? 

The answer, as we will show, is that amplification  is not necessarily linear. That is, in the Heisenberg picture, the transformation of the bosonic annihilation operator can be nonlinear while still preserving the bosonic commutation relation.
And, perhaps surprisingly, that way of amplifying can decrease the amount of noise added.

\section{Nonlinear amplification}The idea is that for detecting single photons it is sufficient  to have an output field whose total  number of excitations is given by $N_{\rm out}=N_{\rm in}+G n_a$ with $n_a$ the number of input photons we would like to detect, and $N_{\rm in}$ the (fluctuating) number of excitations initially present in the output mode, which is {\em not} amplified.
A physically allowed but highly idealized unitary transformation that accomplishes this is easiest 
written down in the Schr\"odinger picture (valid for any $n$ \footnote{The transformation (\ref{simple}) can be realized only when $M\geq Gn$. 
There is always such a restriction on amplification relations; the energy transferred to reservoir 2 must come from somewhere.}, even though in practice we will be interested mainly in small values of $n$, say, $n=0,1,2$) as
\bea\label{simple}
\ket{n}_a \ket{M}_1\ket{N}_2 \longmapsto
\ket{n}_a\ket{M-Gn}_1\ket{N+Gn}_2. 
\eea
All states here are number (Fock) states of bosonic modes.
The transformation involves two energy reservoirs: energy is transferred from the first  reservoir to the second with the amount of energy transferred determined by the number $n$ of input photons in mode $a$ (with nothing happening at all when $n=0$).
  The assumption is that excitations of the two reservoirs have identical energies, $\hbar\omega'$, such that energy is conserved. The input mode can have any frequency $\omega$.
The second reservoir ideally starts out with $N=0$ excitations---corresponding to the zero temperature limit---such that in the end it would contain exactly $Gn$ excitations if the input field contained $n$ photons. Clearly, this ideal transformation would represent perfect (noiseless) amplification of a photon number state (and $G$ will have to be an integer for this to work).

  \begin{figure}[t] 
	\includegraphics[width=.8\linewidth]{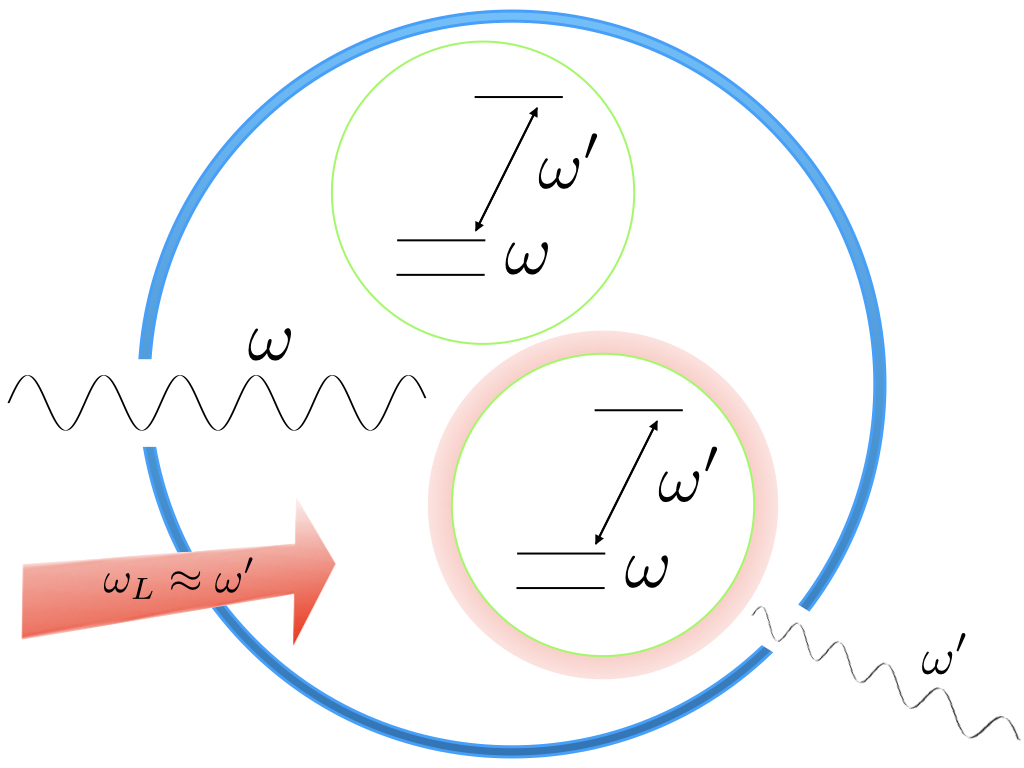} 
	\caption{An input photon with frequency $\omega$ undergoes amplification into a macroscopic signal via electron-shelving \cite{dehmelt1975,wineland1980,bergquist1986}: when an on-resonance photon is absorbed, an atom (modeled here as a three-level system) enters the first excited state and a laser tuned to the second transition frequency $\omega_L\approx \omega'$ induces fluorescence. If there are multiple input photons, they are absorbed by multiple atoms and the fluorescence signal is increased proportionally. The number of fluorescence modes may be reduced by using a high-Q cavity so that amplification moves towards the ideal transformation given in Eq.~\ref{simple}.}
	\label{idealamp}
\end{figure}

Now we wish to describe this ideal process in the Heisenberg picture so as to make a direct comparison with Eq.~(\ref{caves}).
In that picture, the ideal transformation that is linear in the number operator for the excitations in the second reservoir should be \footnote{\label{construction}
The construction of the transformation (\ref{simpleb}) (and transformations including higher powers of the input photon
number operator) is highly constrained by two conditions: that the spectrum of the operator-representation
be  $\mathbb{N}^0$ (the natural numbers including zero) and that the
commutator be preserved.  From these, we need \emph{at least} one term with a number operator on the right in (\ref{simpleb}) with a prefactor of one. Additional reservoirs with arbitrary prefactors are allowed but they will carry additional noise and decrease the SNR.}
\be\label{simpleb}
\hbd_{\rm out}\hb_{\rm out}=\hbd_{\rm in}\hb_{\rm in}+G\had_{\rm in}\ha_{\rm in}.
\ee
We are now going to do three things: (A) we will construct expressions for $\hbd_{\rm out}$ and $\hb_{\rm out}$ 
such that (\ref{simpleb}) is reproduced and such that their commutator 
$[\hbd_{\rm out},\hb_{\rm out}]=\openone$;  (B)
we will add non-ideal features that make the model more realistic, and (C) we will include fluctuations in the initial number of excitations in the $b$ mode and calculate the signal-to-noise ratio (SNR) for the final number of excitations in the $b$ mode, both for the ideal limit and the more realistic models. A comparison with linear amplification will then show how nonlinear amplification improves upon the former.

To start with part of task (B), we adjust the idealized Schr\"{o}dinger picture to get rid of two features that make the process (\ref{simple}) obviously inapplicable to real detectors, but such that the Heisenberg picture (\ref{simpleb}) is still valid. 
First note that the $n$ photons in the process (\ref{simple}) are not destroyed, whereas in a standard detector they are. We fix that by introducing another quantum system $S$ with a continuum of energies $E$ that can absorb the energy $n\hbar\omega$ of the $n$ photons. This modifies (\ref{simple}) by adding a step
\be
\ket{n}_a\ket{E}_S
\longmapsto\ket{0}_a\ket{E+n\hbar\omega}_S.
\ee
Since this extra step does not affect the state of reservoir 2, the crucial equation (\ref{simpleb}) stays the same. The second change concerns phase: in the Schr\"{o}dinger picture we can insert random phase factors $\exp(i\phiclub)$ on the right-hand side of (\ref{simple}). This makes the amplification process irreversible (as any amplification process in a real detector is) and it destroys superpositions of different number states 
(e.g., coherent states will not be coherently amplified). It destroys any entanglement between the different modes as well \footnote{Phase randomization is necessary for optimal amplification and measurement of photon number due to number-phase uncertainty.}.

For task (A) we would like to use the polar decompositions of the creation and annihilation operators. That is, in analogy to the polar decomposition of a complex number, $z=\exp(i\phi)\sqrt{|z|^2}$, we would like to write
\bea\label{bb}
\hat{b}_{{\rm out}}&= \hat{S}\,\sqrt{(\hbd\hb)_{{\rm in}}+G(\had\ha)_{{\rm in}}},
\eea
where $\hat{S}$ is a unitary operator written in the suggestive form $\exp(i\hat{\phi})$ for some hermitian operator $\hat{\phi}$.
In a finite-dimensional Hilbert space of dimension $s+1$ there is no problem defining $\hat{S}$: it is a  shift operator that acts on number states $\ket{N}$ of the bosonic mode as
\bea\label{S}
\hat{S}\ket{N}=e^{i\phiclub}\ket{N-1}\,\,\,{\rm for}\,\,s \geq N>0,
\eea
with $\hat{S}\ket{0}=\ket{s}$ and $\phiclub$ the random phase we introduced earlier.  Since Fock space is  infinite-dimensional, we 
use the Pegg-Barnett trick \cite{pegg1989} 
of truncating the Hilbert space at a high excitation number $s$ and only in the end (when calculating physical quantities) taking the limit $s\rightarrow\infty$. 
It is easy to verify that the relation (\ref{bb})
yields the commutator $[\hb_{{\rm out}},\hbd_{{\rm out}}]=\openone_{{\rm in}} - (s+1) \ket{s}\bra{s}$, in which the extra Pegg-Barnett term won't contribute to any physical quantity, while ensuring a traceless commutator, necessary in finite dimensions \footnote{Note the dimension of both input and output mode Hilbert spaces is $s+1$; they necessarily match in the Heisenberg picture.}.  

The nonlinear equation (\ref{bb}) does not seem to have appeared in the large literature on bosonic amplification (for a review, see, e.g., \cite{clerk2010}). 
Refs.~\cite{yuen1986,ho1994,yuen1996} did discuss photon-number amplifiers (especially in the high-photon number limit) decades ago, but no attempt was made there to find commutator-preserving operator equations.

\section{More realistic models for amplification}\label{Models} Continuing with task (B), 
in a more realistic description the reservoirs consist of many modes.
So, instead of having just one bosonic output mode $b$ we really should describe many output reservoir modes.
For example, we may have $G$ modes $b_k$ [recall $G$ is an integer now] each one of which satisfies
\bea\label{bk}
\hat{b}_{k\,{\rm out}}&=\hat{S}_k\,\sqrt{(\hbd\hb)_{k\,{\rm in}}+(\had\ha)_{{\rm in}}},\,\,\,\,\,k=1\ldots G.
\eea
Here the macroscopic signal monitored and analyzed consists of the {\em sum} of all detected excitations (since each mode by itself contains just a microscopic number of excitations we cannot simply assume to be able to count those individual numbers: then we would not need amplification at all!). That is, we consider as our macroscopic output signal
\bea\label{Iout}
\hat{I}_{{\rm out}}=
\sum_{k=1}^G (\hbd\hb)_{k\,{\rm out}}
=\sum_{k=1}^G (\hbd\hb)_{k\,{\rm in}} + G (\had\ha)_{{\rm in}}.
\eea
Another extension is to ``avalanche" photodetection where one small-scale amplification event triggers the next and the process repeats, giving rise to a macroscopic signal. Iterating the transformation (\ref{simpleb}) of single mode amplification $N$ times with a gain factor $g$ in each step gives a total gain factor $G=g^N$ and an input-output relation 
\be\label{simplebn}
(\hbd\hb)_{\rm N \,out}=\sum_{k=1}^{N}g^{N-k} (\hbd\hb)_{k\,{\rm in}} + G (\had\ha)_{{\rm in}}
\ee 
where mode $\hat{b}_k$ here contains the output of the $k$th amplification step, and the last mode $b_N$ contains the signal. 

Another extension, relevant for $n>1$, describes multiplexing: the idea is that $n$ photons are most conveniently detected by $n$ detectors that each detect one (and only one) photon, along the lines of \cite{nehra2017,yu2018}. We will not describe this model in any detail, except to state that
amplification would in that case be described by
$Gn$ modes, each containing exactly one extra excitation. 

Lastly, we combine both multi-mode and multi-step extensions above by repeating the process in (\ref{bk}) and (\ref{Iout}) of amplification into several ($g$) modes $N$ times, again with a total gain factor defined $G=g^n$ and an input-output relation for the macroscopic signal
\bea\label{Ioutn}
\\
\hat{I}_{{\rm out}} = \sum_{k_N=1}^G (\hbd\hb)_{k_N\,{\rm out}}
= \sum\limits_{n=1}^{N}\sum\limits_{k_n=1}^{g^n} (\hbd\hb)_{k_n\,{\rm in}} + G(\had\ha)_{{\rm in}}\nonumber
\eea
where the mode $\hat{b}_{k_n}$ is the $k_n$th mode in the $n$th step.

Note that in our nonlinear amplification models the amplified signal ends up in different bosonic mode(s): indeed, a photodetector typically converts the input signal (light) to an output signal of a physically different type, e.g., electron-hole pairs (which may sometimes be approximated as composite bosons; see also \cite{keldysh1968,devoret2000,laikhtman2007,combescot2007}).

\section{Number fluctuations}We turn to task (C) and calculate the noise in photon number introduced by the amplification process and by the coupling to reservoirs. 
For the reservoir we monitor, we write 
\bea\label{bvar}
\expect{(\hbd \hb)_{\rm{in}}}=\overline{n}_b;\,\expect{(\hbd \hb)^2_{\rm{in}}}=\overline{n}^2_b + \Delta n_b^2
\eea 
and make no further assumptions about its initial state.

We assume that there is some (unknown) number of photons in the input mode $a$ that we want to measure. We thus consider input states that are diagonal in the photon number basis, with some nonzero photon number fluctuations $\Delta n_a$. 
(Thanks to the randomized phase assumption we can use this assumption without loss of generality for our nonlinear models.)
So, we write
\bea\label{avar}
\expect{(\had\ha)_{\rm{in}}}=\overline{n}_a;\,\expect{(\had\ha)^2_{\rm{in}}}=\overline{n}^2_a +\Delta n_a^2.
\eea 
In the following we always assume the initial states of modes $a$ and $b$ to be independent, such that
\be
\expect{f(\ha,\had)g(\hb,\hbd)}=\expect{f(\ha,\had)}
\expect{g(\hb,\hbd)}
\ee
for any functions $f$ and $g$.

 For linear phase-insensitive amplification (\ref{caves}), we find the following variance in the number of excitations in the amplified signal:
\bea
\label{cavesvar}
\sigma_{(\had\ha)_{\rm out}}^2 &=&G^2\Delta n_a^2 + (G-1)^2 \Delta n_b^2 +\nonumber\\
&& G(G-1) (2 \overline{n}_a\,\overline{n}_b + \overline{n}_a+\overline{n}_b + 1).
\eea
Not only are the fluctuations in the auxiliary mode $b$ amplified [second term in (\ref{cavesvar})], there is inherent noise from the amplification process itself even if $\Delta n_b^2=0$ [the second line is strictly positive for $G>1$]. 

We should also consider linear phase-sensitive amplification \cite{caves1982}, described by 
\bea\label{phase}
\ha_{{\rm out}}=\sqrt{G}\ha_{{\rm in}}+\sqrt{G-1}\had_{{\rm in}}.
\eea
Here, compared to (\ref{caves}) the $\hbd$ term is replaced by the $\had$ term, such that the commutator $[\ha_{{\rm out}},\had_{{\rm out}}]$ is still preserved.
This gives  a variance
\bea
\label{phasevar}
\!\!\sigma_{(\had\ha)_{\rm out}}^2 &=& (6G(G-1) + 1) \Delta n_a^2 +\nonumber\\
&& 2G(G-1) (\overline{n}^2_a+\overline{n}_a+1).
\eea 
There is  again extra amplification noise  for $G>1$ [the second line], much like what we found for phase-insensitive amplification.

We compare these two results for linear amplification to the result for the nonlinear amplification process described by (\ref{bb}). The variance in excitation number is
\bea\label{bbvar}
\sigma_{(\hbd\hb)_{\rm out}}^2 &=\Delta n_b^2 + G^2 \Delta n^2_a.
\eea 
Here the number fluctuations in the auxiliary mode are {\em not} amplified and there is no additional amplification noise either. Already we are able to see that the scheme of amplification into a single mode is optimal; any transformation that would reduce the prefactor of $\Delta n_b^2$ in (\ref{bbvar}) below unity would fail to realize a well-behaved annihilation operator (for details, see again \footnotemark[3])! 

For nonlinear amplification into many modes described by the more realistic model equations (\ref{bk}) 
and (\ref{Iout}), we find 
\bea\label{MNvar}
\sigma_{\hat{I}_{\rm out}}^2 &=G\Delta n_b^2 + G^2 \Delta n^2_a,
\eea 
where for simplicity we assumed all reservoir modes to be independent with the same number fluctuations. This shows amplifying according to
(\ref{bk}) is suboptimal; even though it still beats both linear amplification limits (\ref{cavesvar})
and  (\ref{phasevar}) the noise in the reservoir modes is still amplified. Similarly, amplification using multiple fermionic degrees of freedom will be sub-optimal; a similar multi-mode description will be necessary \footnote{One way around this limitation is for the incident photons to only interact with a single symmetrized collective degree of freedom of many fermions, on which a measurement is then made. In this idealized case, this collective degree of freedom plays the role of a single bosonic mode and amplification could still be described by (\ref{bb}) and photon number amplification is improved past the limit for linear fermionic amplification \cite{yurke2004}.}.

Defining the total gain $g^N=G$ with $N$ the number of steps, we find for our multi-step models that 
\bea\label{iter}
\sigma_{(\hbd\hb)_{\rm out}}^2=\frac{G^2-1}{g^2-1}\Delta n^2_b+G^2\Delta n^2_a
\eea
for amplification of $g$ excitations into a single mode and 
\bea\label{itermult}
\sigma_{\hat{I}_{\rm out}}^2=G\frac{G-1}{g-1}\Delta n^2_b+G^2\Delta n^2_a
\eea
for amplification of a single excitation into $g$ modes. 

\section{Signal-to-noise ratios}We can now write down explicit tradeoff relations between amplification and number fluctuations in terms of signal-to-noise ratios for all types of amplification discussed here, for the case where the number of input photons is {\em fixed} to be $n_a$  (and so $\Delta n_a=0$).
Using the standard signal-to-noise ratio as the number of excitations in the amplified mode minus the background, divided by the standard deviation in the number of excitations, we get 
\bea\label{all}
\textnormal{SNR}_{{\rm PhaseInsensitive}} &\leq&\frac{G}{G-1}\frac{n_a}{\Delta n_b}\label{PhaseInsensitive}\\
\textnormal{SNR}_{{\rm PhaseSensitive}}  &\leq&\frac{2G - 1}{\sqrt{2G(G-1)}} n_a\label{PhaseSensitive}\\
\textnormal{SNR}_{{\rm SingleMode}}&=& \frac{Gn_a}{\Delta n_b}\label{SM}\\
\textnormal{SNR}_{{\rm GModes}}&=&\frac{Gn_a}{\sqrt{G}\Delta n_b}=\frac{\sqrt{G}n_a}{\Delta n_b}.\label{GM}
\eea
The linear amplification mechanisms have increasingly worse signal-to-noise ratios as $G$ increases \footnote{The signal-to-noise ratios (\ref{PhaseInsensitive}) and (\ref{PhaseSensitive}) for linear amplification become infinite at $G=1$ simply because there is no noise when both $G=1$ and $\Delta n_a = 0$.}, albeit saturating in the limit $G\rightarrow\infty$.
In contrast, the signal-to-noise ratios for the nonlinear amplification mechanisms improve with increasing $G$, with amplification into a single-mode performing best \footnote{We find the linear dependence on $G$ resulting from single-shot single-mode amplification holds for transformations describing higher-order amplification of photon number operator, again subject to the constraints of [13].}.


Similarly, we consider multi-step amplification models
\bea
\textnormal{SNR}_{{\rm MultiStepSingleMode}}&=&\frac{G\sqrt{g^2-1}n_a}{\sqrt{G^2-1}\Delta n_b}\label{ISM}\\
\textnormal{SNR}_{{\rm MultiStepMultiMode}}&=& \frac{\sqrt{G(g-1)}n_a}{\sqrt{G-1}\Delta n_b}.\label{IMM} 
\eea 
These intermediate noise limits fill in the space between the optimal SNR (\ref{SM}) and linear amplification. (Indeed, multi-step multi-mode amplification performs slightly \emph{worse} than both linear mechanisms for $g=2$!)


  


\section{Single-photon pre-amplification}While our paper focuses on the amplification part of the photo detection process,  we very briefly consider the pre-amplification process now. We certainly cannot treat that part in full generality here and we adopt several simplifications in order to arrive at an important result concerning the suppression of thermal noise. First, we assume that we can decouple the amplification stage from the pre-amplification filtering  [by having an irreversible step in between the two] such that filtering does not interfere negatively with the absorption/transduction part \cite{young2018}. We then focus on just the time/frequency degree of one incoming photon \footnote{
Photon-number resolved photo detection can be achieved
by multiplexing an $n$-photon signal to many ($N\gg n$) single photon detectors \cite{nehra2017}, each satisfying (\ref{transfer}) independently. However, this means an additional noise mode will be added with each splitting of the signal, decreasing the integrated signal-to-noise ratio.  To avoid added noise a non-linear multi-photon filtering process could be used, but for this a full S-matrix treatment must be used, see Refs. ~\cite{fan2010,caneva2015,xu2015}.}.  A single absorber with some resonance frequency $\omega_0$ able to absorb that single photon will act as a frequency filter. If the pre-amplification filtering is passive (easy to implement,  but we certainly can go beyond this \footnote{See, for example, Ref.~\cite{raymer2010}. The result is that, instead
of certain frequencies, it is certain spectral ``Schmidt modes'' that are detected perfectly.}) and unitary (i.e., lossless: we consider this because we are interested in the fundamental limits of photo detection. Internal losses only degrade performance.), then frequency filtering is described by the linear transformation 
\be\label{transfer}
\ha_{{\rm out}}(\omega)=T(\omega) \ha_{{\rm in}}(\omega)+R(\omega)\hat{c}_{{\rm in}}(\omega)
\ee 
where $c_{{\rm in}}(\omega)$ is yet another internal bosonic detector mode at the same frequency as the input mode \cite{Propp}. Here $T(\omega)$ and $R(\omega)$ are ``transmission'' and ``reflection'' coefficients which
satisfy 
$|T(\omega)|^2+|R(\omega)|^2=1$ and which are determined by the resonance structures internal to the photodetector. 
The amplification process that follows the initial absorption of the photon energy is applied to
the operator $\hat{a}_{\rm out}(\omega)$ of Eq. (\ref{transfer}), so that (explicitly displaying the different frequencies of the modes now) ideal amplification (single-mode and single-shot) is described 
\be\label{simplec}
\!\hbd_{\rm out}(\omega')\hb_{\rm out}(\omega')=\hbd_{\rm in}(\omega')\hb_{\rm in}(\omega')+G\had_{\rm out}(\omega)\ha_{\rm out}(\omega).
\ee
This makes rigorous the idea that one {\em can} amplify at any frequency, enabling the mantra that one {\em should} amplify at high (optical) frequencies \cite{Dowling}. Namely, thermal fluctuations at a frequency $\omega'$ may be suppressed by choosing the reservoir mode frequency $\omega'$ such that
$\hbar\omega'\gg kT$. This suppression is exponential: $\Delta n_b^2\propto\bar{n}_b\propto \exp(-\hbar\omega'/kT)$. 
Note that number fluctuations in the internal mode $c_{{\rm in}}(\omega)$ at the input frequency will be amplified by the subsequent amplification process.
However, one can in principle construct ideal detectors for light with a particular frequency $\omega_0$ \cite{young2018}, such that $|T(\omega_0)|=1$ and hence $R(\omega_0)=0$ \cite{Propp}, avoiding internally generated dark counts at that particular frequency. 
If instead of a single spectral mode a small range of frequencies is amplified with differing probabilities, matching the amplification spectrum to the filtering spectrum is sufficient for reducing internally generated dark counts, as we will discuss in more detail in work in preparation \cite{Propp2}.

\section{Further applications}The models for amplification considered here apply to other types of quantum measurement as well.
For example, electron-shelving \cite{dehmelt1975,wineland1980,bergquist1986} is a well-known method to perform atomic state measurements. Here one particular atomic  state (e.g., one of the hyperfine ground states of an ion) is coupled resonantly to a higher-lying excited state which can then decay back by fluorescence only to that same ground state.
A laser tuned to that transition can then induce the atom to emit a macroscopic amount (visible by eye) of fluorescent light.
In the language accompanying Eq.~(\ref{simple}), the laser beam forms the first reservoir, while the second reservoir consists of vacuum modes that are filled with fluorescent light as described by (\ref{bk}). The gain factor $G$ (the number of fluorescence photons) is determined by the ratio of Einstein's coefficients for spontaneous and stimulated emission and the total integration time.  
By placing the atom/ion inside a high-Q optical resonator (with resonant frequency $\omega'$) we would reduce the number of output modes and thereby get closer to the optimum. 
The idea of placing a detector inside a resonant cavity is, of course, not new \cite{unlu1995}, but that idea is usually associated with increasing the coupling to light. Although we do have that effect as well, the main purpose here is to reduce the number of output modes, and thereby increase the SNR (Fig.~\ref{idealamp}).

In Ref.~\cite{yuen1986} a transformation similar to (\ref{simple}) is given, with $n$ photons being converted to $Gn$ photons in a single mode directly. Though this transformation is unphysical (there is no way to preserve the commutator), a SNR is calculated that increases linearly with $G$ like our Eq. (\ref{SM}). However, the SNR found in \cite{yuen1986} diverges for a photo detector with unit efficiency, which is not the case once fluctuations in the reservoir mode are properly taken into account as our results clarify.

In Ref.~\cite{yang2019} an interesting model for amplification is constructed that makes use of a first-order phase transition for a collection of $N$ interacting spin-1/2 particles. These spins are coupled both to an input photon and to an output bosonic mode. The SNR (as we define it here) for that model
scales as $\sqrt{N}$ while the gain $G$ of that model is linear in $N$. Thus, the SNR scales with $\sqrt{G}$ just as our Eq.~(\ref{GM}): the number of spins in Ref.~\cite{yang2019}'s model plays a similar role as our number of amplification modes.


\section{Conclusions}We discussed various linear and nonlinear  amplification schemes for bosonic modes. For detecting few photons, we found that the latter add considerably less noise, leading to better signal-to-noise ratios, as exemplified in Eqs.~(\ref{all})--(\ref{IMM}). Unlike for linear amplification, number fluctuations in internal detector modes are not amplified, while the number of photons that we want to detect {\em is} amplified. All amplification schemes explicitly preserve the bosonic commutation relations.

While amplification into a single-mode may not be feasible in practice, it provides the fundamental lower limit to noise in photon-number measurements across amplification mechanisms. In practice,
one may have many output modes and thus may find a SNR  closer to Eq. (\ref{GM}), which is worse by a factor of $\sqrt{G}$ than the fundamental limit (but still better by a factor of $\sqrt{G}$ than linear amplification), or one may have multiple amplification steps (\ref{ISM}), or both (\ref{IMM}). To test this,
we suggest that measurement of the gain dependence of the SNR for a given photo detector should provide a rough but useful indication of the underlying amplification mechanism.

\vspace{1em}

This work is supported by funding from
DARPA under
Contract No. W911NF-17-1-0267.
We thank Joseph Altepeter and Sae Woo Nam for their useful comments on this project.

\bibliography{amplification3}

\end{document}